\documentclass[%
preprint,
 amsmath,amssymb,
aps,
prb,
showkeys,
endfloats*,
multicol,
]{revtex4-1}

\usepackage{graphicx}
\usepackage{dcolumn}
\usepackage{bm}
\usepackage{color}
\usepackage{txfonts}

\begin{document}
\title{Density functional theory calculations for investigation of atomic structures of 4H-SiC/SiO$_2$ interface after NO annealing}
\author{Naoki Komatsu}
\affiliation{Department of Electrical and Electronic Engineering, Graduate School of Engineering, Kobe University, Nada, Kobe, 657-8501 Japan}
\author{Mizuho Ohmoto}
\affiliation{Department of Electrical and Electronic Engineering, Graduate School of Engineering, Kobe University, Nada, Kobe, 657-8501 Japan}
\author{Mitsuharu Uemoto}
\affiliation{Department of Electrical and Electronic Engineering, Graduate School of Engineering, Kobe University, Nada, Kobe, 657-8501 Japan}
\author{Tomoya Ono}
\email{t.ono@eedept.kobe-u.ac.jp}
\affiliation{Department of Electrical and Electronic Engineering, Graduate School of Engineering, Kobe University, Nada, Kobe, 657-8501 Japan}

\date{\today}

\begin{abstract}
We propose the atomic structures of the 4H-SiC/SiO$_2$ interface for the $a$, $m$, C, and Si faces after NO annealing. Our proposed structures preferentially form at the topmost layers of the SiC side of the interface, which agrees with the experimental finding of secondary-ion mass spectrometry, that is, the N atoms accumulate at the interface. In addition, the areal N-atom density is on the order of 10$^{14}$ atom/cm$^2$ for each plane, which is also consistent with the experimental result. Moreover, the electronic structure of the interface after NO annealing, in which the CO bonds are removed and the nitride layer only at the interface is inserted, is free from gap states, although some interface models before NO annealing include the gap states arising from the CO bonds near the valence band edge of the bandgap. Our results imply that NO annealing can contribute to the reduction in the density of interface defects by forming the nitride layer.
\end{abstract}

\maketitle

\section{Introduction}
Silicon carbide (SiC) attracts attention because it is one of the most promising wide-bandgap semiconductors for developing next-generation switching devices operating in high-power and high-frequency applications.\cite{JpnJApplPhys_45_7565, JpnJApplPhys_54_040103, ApplPhysExp_13_120101} However, the potential of 4H-SiC has not been fully utilized owing to the low carrier mobility of 4H-SiC-based metal-oxide-semiconductor field-effect transistors, which is caused by defects at the 4H-SiC/SiO$_2$ interface. To reduce the density of interface defects and increase the channel mobilities, post-oxidation annealing using N$_2$O, NO, or N$_2$ gas has been proposed.\cite{ApplPhysLett_70_2028, ApplPhysLett_76_1713, JApplPhys_112_024520, Energies_12_2310, JpnJApplPhys_58_SBBD04} In previous studies for the experimental characterizations of the annealed SiC/SiO$_2$ interface, the incorporated N atoms were fixed at the C atom sites of the SiC substrate side, which were chemically bonded with Si atoms.\cite{JSurfSciNanotechnol_15_109, ApplPhysLett_99_182111} The presence of a high-N-atom-density layer, which is called the nitride layer hereafter, was observed by the secondary-ion mass spectrometry of the $a$ face (1$\bar{1}$00), $m$ face (11$\bar{2}$0), Si face (0001), and C face (0001). Since the N atom density is on the order of 10$^{14} - 10^{15}$ atom/cm$^2$ for each crystal plane,\cite{JSurfSciNanotechnol_15_109, JApplPhys_97_074902} it is expected that most of the C atoms at the interface are substituted by N atoms. On the basis of their low-energy electron diffraction measurements, Shirasawa and coworkers proposed a model of the N-annealed 6H-SiC/SiO$_2$ interface, which constitutes epitaxially stacked SiO$_2$ and Si$_3$N$_2$ monolayers (SiON) on SiC.\cite{PhysRevLett_98_136105, PhysRevB_79_241301} A similar structure was also determined for 4H-SiC using the simulated X-ray absorption spectroscopy spectrum.\cite{JSynchrotronRad_26_462} Moreover, some experiments reported that the N atom density after annealing is different depending on the crystal planes. The N nuclear reaction analysis of the 4H-SiC/SiO$_2$ interface showed that the concentration of N atoms on the $a$ face is more than twice that on the Si face. Hamada {\it et al.} \cite{JSurfSciNanotechnol_15_109} reported that the $m$ face has nearly the same concentration of N atoms as the $a$ face. Regardless of the annealing temperature, N-atom incorporation occurs in the order of $a$ $\sim$ $m$ $\ll$ Si, excluding the C face. Although the anisotropy mechanism has been investigated so far, the atomic-scale structures of the interface after annealing have been unclear for arbitrary crystal planes. In our previous study,\cite{JPhysSocJpn_90_124713} we proposed the interface atomic structures in which N atoms are incorporated by replacing C atoms, and we investigated the total energies and electronic structures of the nitride layers by density functional theory (DFT) calculation.\cite{PhysRev_136_B864} It was found that the incorporation at the $k$ site on the $a$ face is energetically most stable, and the areal density of N atoms in our model ($\sim 10^{14}$ atom/cm$^2$) agrees well with experimental results. However, the previous study employed the models where the nitride layers form in a SiC bulk and the interaction between the substrate and oxides is not considered.

In this study, we employ the interface models in which the atomic structures include oxides to investigate the effects of the interaction between the substrate and oxides. Universal atomic-scale models describing 4H-SiC with a high-N-atom-density layer in arbitrary crystal planes are proposed. To explore the most preferable crystal plane of 4H-SiC for N-atom incorporation, we study the stability and electronic states of the nitride layers on a 4H-SiC surface. We find that the nitride layers grow along the $a$ face, which is consistent with the conclusion obtained using the bulk model,\cite{JPhysSocJpn_90_124713} and NO bonds are hardly generated at the interface. Any defect states in the bandgap of 4H-SiC are not generated in our structure. The formation energy of the nitride layers at the topmost layer of the interface is smaller than that at the second layer, indicating that the N atoms accumulate at the interface.

The rest of this paper is organized as follows. Section~\ref{sec:Method} is devoted to the description of the methods used in this work. In Sec.~\ref{sec:Results and discussion}, we propose the interface atomic structures after N-atom incorporation and consider the formation energies of the proposed structures. The conclusions are drawn in Sec.~\ref{sec:Conclusion}.

\begin{figure}[htb]
\begin{center}
\includegraphics{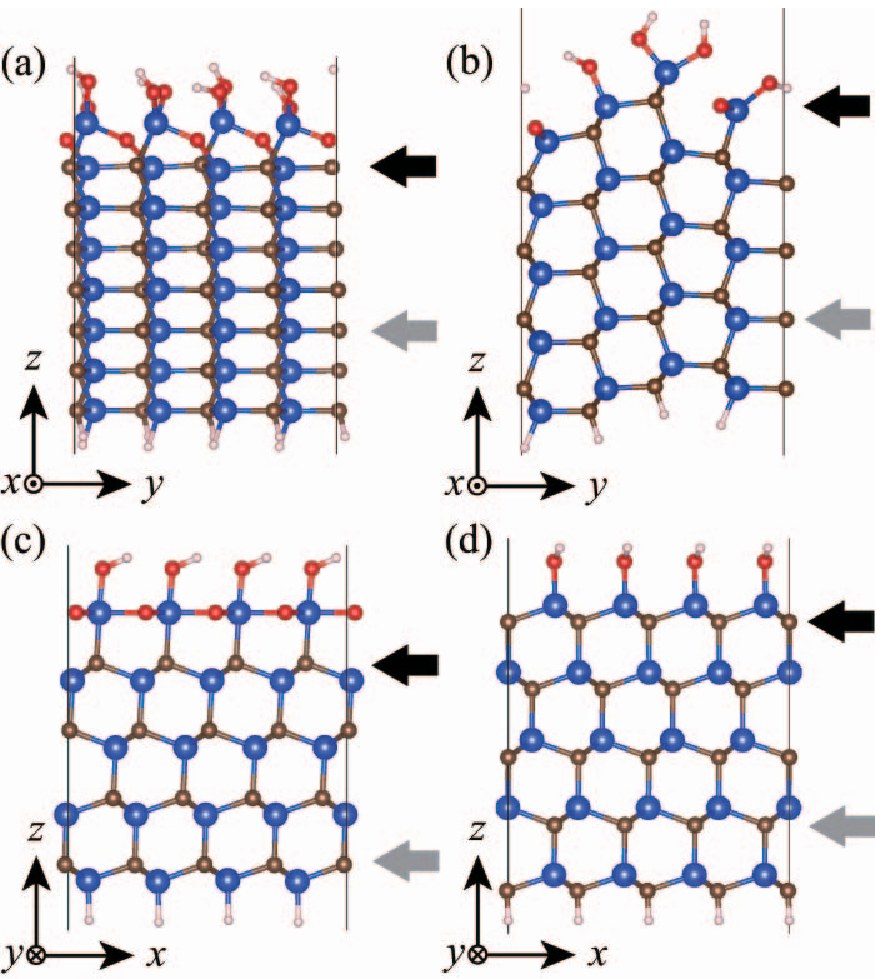}
\caption{Interface atomic structures without CO bonds for (a) $a$-, (b) $m$-, (c) C-, and (d) Si-face models before N-atom incorporation. Blue, brown, red, and gray spheres are Si, C, O, and H atoms, respectively. Black lines are the boundaries of the supercell.}
\label{fig:model1}
\end{center}
\end{figure}

\begin{figure}[htb]
\begin{center}
\includegraphics{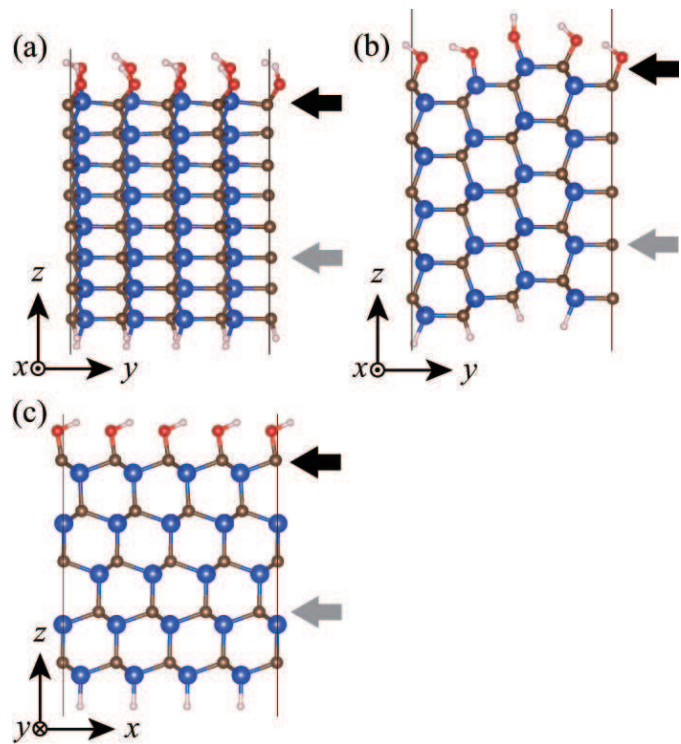}
\caption{Interface atomic structures with CO bonds for (a) $a$-, (b) $m$-, and (c) C-face models before N-atom incorporation. The meanings of the symbols are the same as those in Fig.~\ref{fig:model1}}
\label{fig:model2}
\end{center}
\end{figure}

\begin{figure}[htb]
\begin{center}
\includegraphics{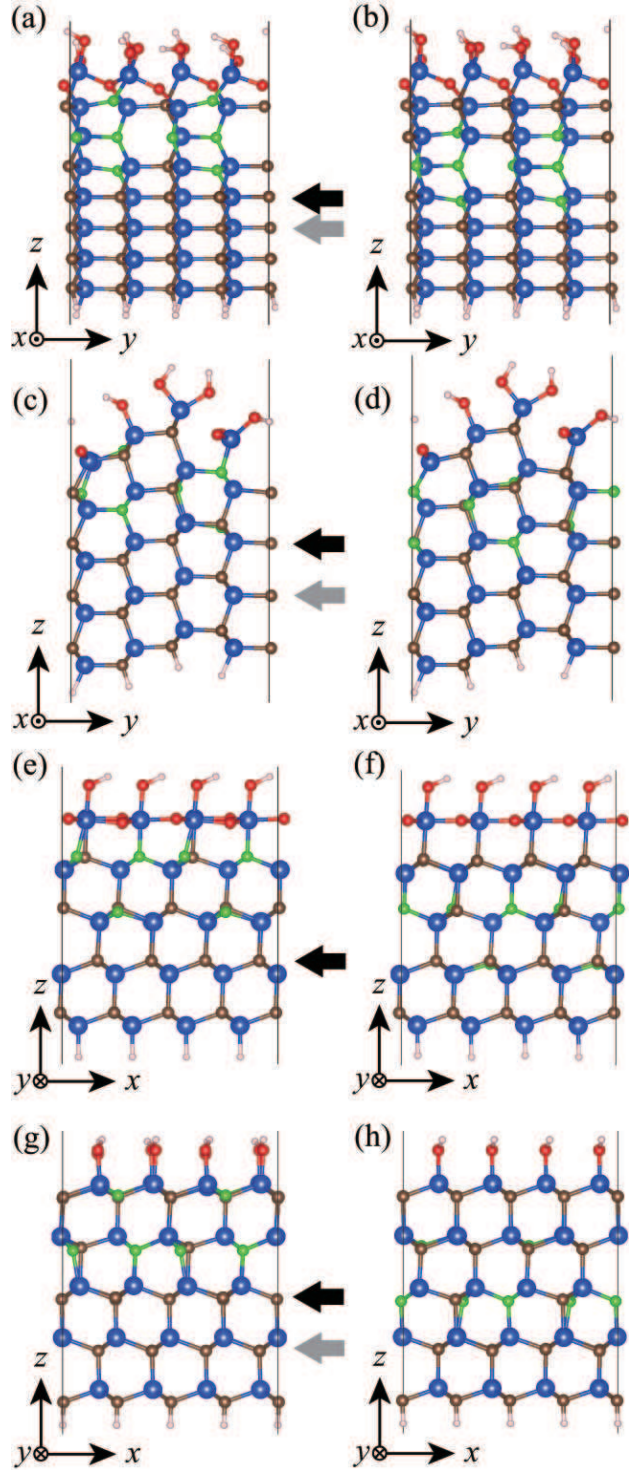}
\caption{Interface atomic structures without CO bonds for (a) [(b)] $a$-, (c) [(d)] $m$-, (e) [(f)] C-, and (g) [(h)] Si-face models with nitride layers at the topmost [second] bilayer. Green spheres are N atoms. The meanings of the other symbols are the same as those in Fig.~\ref{fig:model1}}
\label{fig:model3}
\end{center}
\end{figure}

\begin{figure}[htb]
\begin{center}
\includegraphics{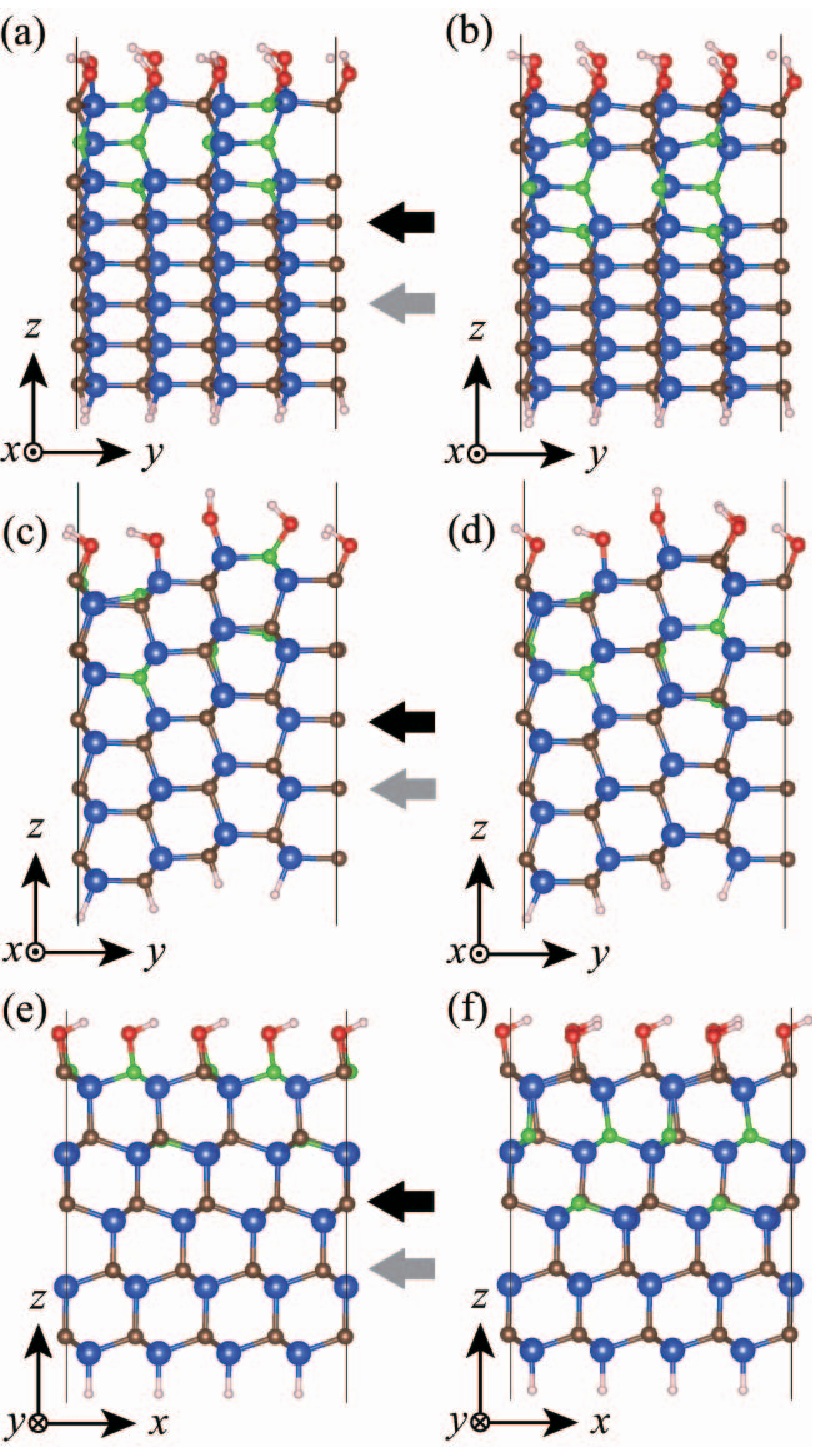}
\caption{Interface atomic structures with CO bonds for (a) [(b)] $a$-, (c) [(d)] $m$-, and (e) [(f)] C-face models with nitride layers at the topmost [second] bilayer. The meanings of the symbols are the same as those in Fig.~\ref{fig:model3}.}
\label{fig:model4}
\end{center}
\end{figure}

\begin{figure}[htb]
\begin{center}
\includegraphics{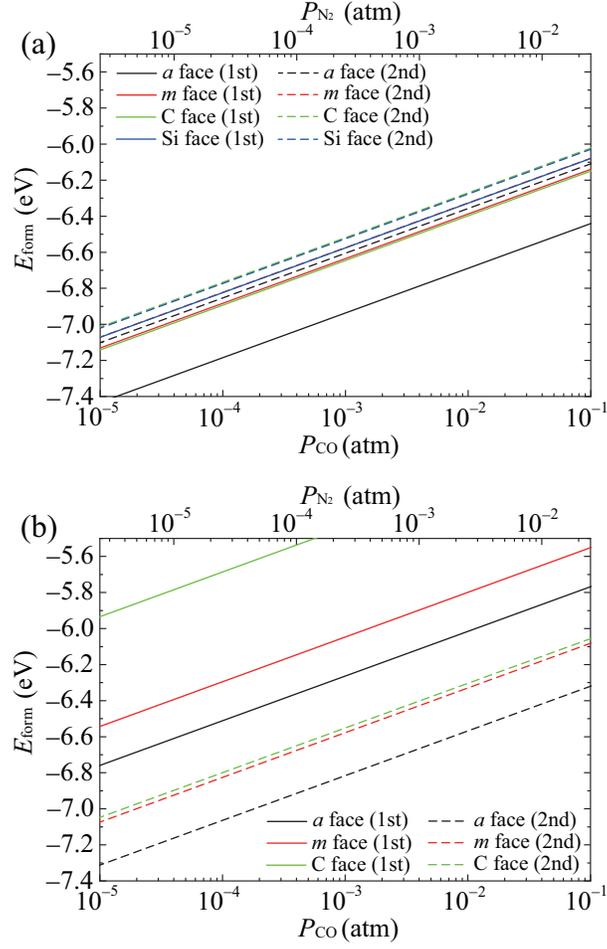}
\caption{Formation energies $E_\mathrm{form}$ defined Eq.~(\ref{eqn:form1}) for interface (a) without and (b) with CO bonds with respect to partial pressures of CO and N$_2$.}
\label{fig:reactionenergy1}
\end{center}
\end{figure}

\begin{figure}[htb]
\begin{center}
\includegraphics{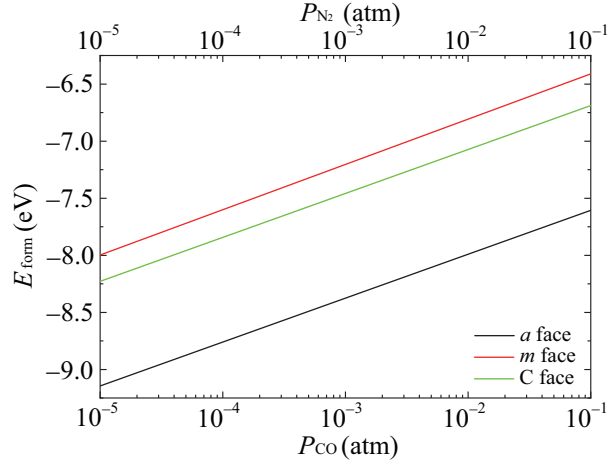}
\caption{Formation energies $E_\mathrm{form}$ defined Eqs.~(\ref{eqn:form2}) and (\ref{eqn:form3}) for interface without CO bonds from that with CO bonds with respect to partial pressures of CO and N$_2$.}
\label{fig:reactionenergy2}
\end{center}
\end{figure}

\begin{figure*}[htb]
\begin{center}
\includegraphics{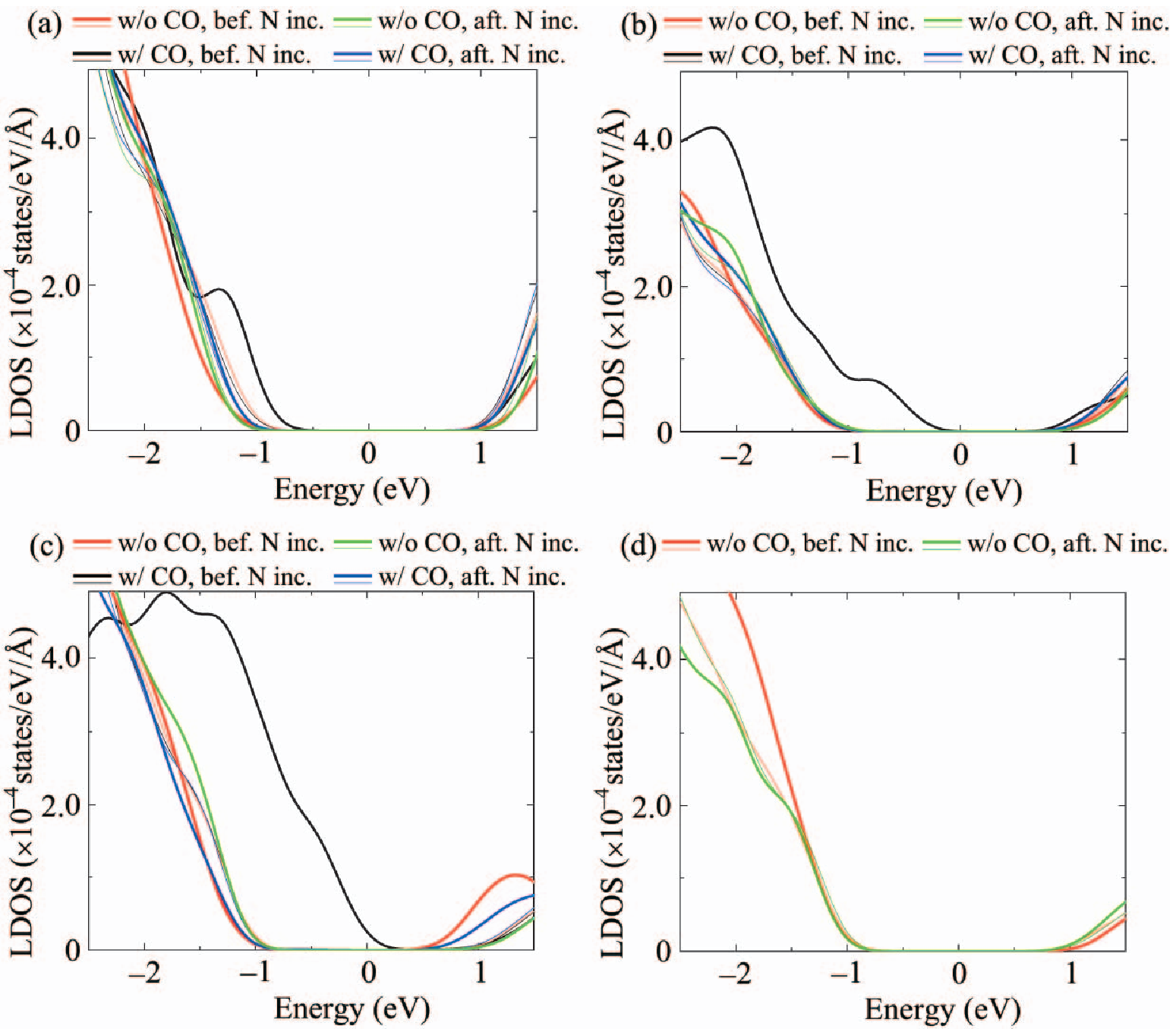}
\caption{LDOS for (a) $a$-, (b) $m$-, (c) C-, and (d) Si-face models. Red and black (green and blue) curves indicate the LDOS before (after) N-atom incorporation. Red and green (black and blue) curves are the LDOS for the interface models without (with) the CO bonds. Thick curves are the LDOS on the planes indicated by black arrows in Figs.~\ref{fig:model1}, \ref{fig:model2}, \ref{fig:model3}, and \ref{fig:model4}, while thin curves are the LDOS on the planes indicated by gray arrows.}
\label{fig:ldos}
\end{center}
\end{figure*}

\section{Method}
\label{sec:Method}
In our model, the surface of the 4H-SiC substrate is terminated with hydroxy groups to imitate the SiO$_2$ layer and the other side of the substrate is terminated with H atoms. We employ rectangular supercells of 10.1 \AA~$\times$ 10.7 \AA~$\times$ 26.3 \AA~ for the $a$-face model, 10.1 \AA~$\times$ 12.3 \AA~$\times$ 27.3 \AA~ for the $m$-face model, and 12.3 \AA~$\times$ 10.7 \AA~$\times$ 26.8 \AA~ for the Si(C)-face model. The direction perpendicular to the surface is taken to be the $z$-axis. At the interface, two types of interface atomic structure are prepared, where CO bonds are present in one type and are absent in the other type. The numbers of atoms for the $a$-face, $m$-face, and Si(C)-face models are 176, 208, and 208, respectively, and the numbers are not affected by the presence of the CO bonds. Figures~\ref{fig:model1} and \ref{fig:model2} show the interface atomic structures before N-atom incorporation without and with the CO bonds, respectively. Although it remains to be clarified whether the inserted N atoms exist at the SiO$_2$ side or the SiC side of the interface, we assume that the N atoms accumulate at the SiC side on the basis of the possibility proposed in the previous experimental studies.\cite{JSurfSciNanotechnol_15_109, JApplPhys_97_074902} Figures~\ref{fig:model3} and \ref{fig:model4} show the interface atomic structures with the nitride layers for the interfaces without and with the CO bonds, respectively. The modification incorporating four Si vacancies (V$_\mathrm{Si}$s) and 16 N atoms at C sites (N$_\mathrm{C}$s) is considered. From the similarity of the SiON layer model reported by Shirasawa and coworkers, \cite{PhysRevLett_98_136105, PhysRevB_79_241301} stable structures without dangling bonds are required. We consider the cases in which V$_\mathrm{Si}$s are arranged parallel to the surface to evaluate the anisotropy of the formation energy of the nitride layer. As reported in our previous paper,\cite{JPhysSocJpn_90_124713} the areal N-atom density is on the order of 10$^{14}$ cm$^{-1}$. Although there are two inequivalent lattice sites of 4H-SiC, i.e., $h$ (hexagonal) and $k$ (quasi-cubic) sites, our previous studies have shown that nitride layers tend to grow at the $k$ site. Thus, we consider the cases in which V$_\mathrm{Si}$s are arranged at the $k$ site. The interfaces with the nitride layer at the second layer are also investigated for all the faces. For the first-principles calculation, we employ the RSPACE code,\cite{PhysRevLett.82.5016, PhysRevB.72.085115, KikujiHirose2005, PhysRevB.82.205115} which uses the real-space finite-difference approach\cite{PhysRevLett_72_001240, PhysRevB_50_011355} within the frameworks of DFT.\cite{PhysRev_136_B864} The local density approximation\cite{CanJPhys.58.1200} of the DFT is used to describe the exchange and correlation effects. The projector-augmented wave method is used for electron--ion interactions.\cite{PhysRevB.50.17953} We adopt 6 $\times$ 6 $\times$ 1 Monkhorst-pack $k$-point meshes including a $\Gamma$-point in the Brillouin zone. The real-space grid spacing is chosen to be $\sim$ 0.21 \AA. The structural optimization is performed until the residual forces are smaller than 0.001 Hartree/Bohr radius.

\section{Results and discussion}
\label{sec:Results and discussion}
The assumed interaction of the SiC substrate and the arriving NO molecule at the interface is expressed as
\begin{equation*}
\mathrm{SiC}^{(\mathrm{w/o} \: \mathrm{mod})} + 24\mathrm{NO} \rightarrow  \mathrm{SiC}^{(\mathrm{w/} \: \mathrm{mod})} + 4\mathrm{SiO}_2 + 16\mathrm{CO} + 4\mathrm{N}_2,
\end{equation*}
with $\mathrm{SiC}^{(\mathrm{w/o} \: \mathrm{mod})}$ ($\mathrm{SiC}^{(\mathrm{w/} \: \mathrm{mod})}$) being the interface without (with) the modification incorporating four V$_\mathrm{Si}$s and 16 N$_\mathrm{C}$s. The formation energy of the above interaction for generating one V$_\mathrm{Si}$ is obtained as
\begin{equation}
\label{eqn:form1}
E_\mathrm{form}= E_\mathrm{total}^{(\mathrm{w/} \: \mathrm{mod})}/4 + E(\mathrm{SiO}_2) + 4\mu_\mathrm{CO} + 2\mu_\mathrm{N} - E_\mathrm{total}^{(\mathrm{w/o} \: \mathrm{mod})}/4 - 6 \mu_\mathrm{NO},
\end{equation}
where $E(\mathrm{SiO}_2)$ is the total energy of a SiO$_2$ unit in a bulk of quartz SiO$_2$, and $E_\mathrm{total}^{(\mathrm{w/o} \: \mathrm{mod})}$ ($E_\mathrm{total}^{(\mathrm{w/} \: \mathrm{mod})}$) represents the total energy of the interface without (with) N-atom incorporation. In addition, $\mu_\mathrm{NO}$, $\mu_\mathrm{CO}$, and $\mu_\mathrm{N}$ are the chemical potentials of a NO molecule, a CO molecule, and N atoms in a N$_2$ molecule, respectively. We set the temperature at 1000 K and the partial pressure of NO gas ($p_\mathrm{NO}$) at 1 atm. The thermochemical parameters are taken from Ref.~\onlinecite{JANAF}. Since the annealing process is not in an equilibrium state, it is not straightforward to determine the partial pressures of CO ($p_\mathrm{CO}$) and N$_2$ ($p_{\mathrm{N}_2}$) gases. Here, $p_\mathrm{CO}$ ($p_{\mathrm{N}_2}$) is varied between $10^{-1}$ and $10^{-5}$ atm ($0.25 \times 10^{-1}$ and $0.25 \times 10^{-5}$ atm).

The formation energies of the nitride layers with respect to $p_\mathrm{CO}$ and $p_{\mathrm{N}_2}$ are shown in Fig.~\ref{fig:reactionenergy1}. The nitride layers growing along the $a$ face are the most stable among those along the $a$, $m$, C, and Si faces. These results are consistent with the conclusion obtained for the nitride layers in bulks; that is, the nitride layer preferentially grows along the $a$ face from the viewpoint of thermodynamics,\cite{JPhysSocJpn_90_124713} and the dependence of the areal N-atom density on the crystal plane reported in Ref.~\onlinecite{JSurfSciNanotechnol_15_109} is caused by the kinetic factor. In the case of the interfaces without the CO bonds, the formation energy of the nitride layer at the topmost layer is smaller than that at the second layer, resulting in the localization of the N atoms at the interface. The difference in the formation energy between the nitride layers at the topmost and second layers is the largest for the $a$ face. This result agrees well with the observation that the formation energy of the nitride layer along the $a$ face is the smallest.

In the case of the interfaces with the CO bonds, compared with the interfaces with the nitride layer at the topmost and second layers, we find that the formation of the nitride layer at the second layer is stable. When the nitride layers are generated at the topmost layer, the C atoms connected to O atoms are replaced by N atoms. However, the formation of NO bonds is not more preferable than that of the SiN bonds. To ensure the stability of the SiN bonds over the NO bonds, we examine the following interactions with isolated molecules.
\begin{eqnarray}
\label{eqn:react1}
\mathrm{CH}_3\mathrm{OH}+\mathrm{NH}_3 &\longrightarrow& \mathrm{NH}_2\mathrm{OH}+\mathrm{CH}_4 \\
\label{eqn:react2}
\mathrm{CH}_3\mathrm{SiH}_3+\mathrm{NH}_3 &\longrightarrow& \mathrm{NH}_2\mathrm{SiH}_3+\mathrm{CH}_4
\end{eqnarray}
The reaction energy is defined as the total energy of the right hand side subtracted by that of the left hand side. The reaction energy of the interaction Eq.~(\ref{eqn:react1}) is 0.44 eV, while that of Eq.~(\ref{eqn:react2}) is $-$0.17 eV. Among Si, C, N, and O atoms, the electronegativity of a Si atom is the smallest and Si atoms are positively charged at the SiC/SiO$_2$ interface. Therefore, the bonds with positively charged Si and negatively charged N atoms are stable, resulting in the stabilization of the atomic structures with the nitride layer at the second layer. The difference in the formation energy between the nitride layers at the topmost and second layers of the C face is significantly larger than those of the other faces. This is because the number of NO bonds generated by the formation of the nitride layer is the largest in the case of the C face.

The interface atomic structure where the nitride layer exists at the second layer contains the transition layer between the SiO$_2$ and nitride layers. It is counterintuitive that a V$_\mathrm{Si}$ is formed directly at the second layer. The first-principles calculation using the nudged energy band method reveals that the energy barrier of the V$_\mathrm{Si}$ from the topmost layer to the second layer along the (0001) direction is 5.0 eV, indicating that the migration of the V$_\mathrm{Si}$ hardly occurs at the interface. Then, we consider the interactions between NO molecules and the interface with the CO bonds, which is the source of the transition layer. The interactions are expressed as
\begin{equation}
\mathrm{SiC}^{(\mathrm{w/} \: \mathrm{CO})} + 16\mathrm{NO} \rightarrow  \mathrm{SiC}^{(\mathrm{w/o} \: \mathrm{CO})} + 8\mathrm{CO} + 8\mathrm{N}_2,
\label{eqn:react3}
\end{equation}
for the $a$ and $m$ faces and 
\begin{equation}
\mathrm{SiC}^{(\mathrm{w/} \: \mathrm{CO})} + 32\mathrm{NO} \rightarrow  \mathrm{SiC}^{(\mathrm{w/o} \: \mathrm{CO})} + 16\mathrm{CO} + 16\mathrm{N}_2,
\label{eqn:react4}
\end{equation}
for the C face, where $\mathrm{SiC}^\mathrm{(w/o \: CO)}$ and $\mathrm{SiC}^\mathrm{(w/ \: CO)}$ are the interfaces without and with the CO bonds shown in Figs.~\ref{fig:model1} and \ref{fig:model2}, respectively. The formation energies of the reaction Eqs.~(\ref{eqn:react3}) and (\ref{eqn:react4}) are defined as
\begin{equation}
\label{eqn:form2}
E_\mathrm{form} = E_\mathrm{total}^{(\mathrm{w/o} \: \mathrm{CO})}/8 + \mu_\mathrm{CO} + 2\mu_\mathrm{N} - E_\mathrm{total}^{(\mathrm{w/} \: \mathrm{CO})}/8 + 2\mu_\mathrm{NO}
\end{equation}
and
\begin{equation}
\label{eqn:form3}
E_\mathrm{form} = E_\mathrm{total}^{(\mathrm{w/o} \: \mathrm{CO})}/16 + \mu_\mathrm{CO} + 2\mu_\mathrm{N} - E_\mathrm{total}^{(\mathrm{w/} \: \mathrm{CO})}/16 + 2\mu_\mathrm{NO},
\end{equation}
respectively. Here, $E_\mathrm{total}^{(\mathrm{w/} \: \mathrm{CO})}$ ($E_\mathrm{total}^{(\mathrm{w/o} \: \mathrm{CO})}$) is the total energy of the interface with (without) the CO bonds and $p_\mathrm{CO}$ and $p_{\mathrm{N}_2}$ are varied between $10^{-1}$ and $10^{-5}$ atm. Figure~\ref{fig:reactionenergy2} shows the formation energy for the elimination of the CO bonds at the interface. We find that all the reactions are exothermic, indicating that the CO bonds are removed before N-atom incorporation, the NO bonds are hardly generated at the interface, and the transition layer is absent even when the CO bonds exist before N-atom incorporation. Note that the NO annealing generates the nitride layers immediately below the SiO$_2$ layers without any transition layers.

Finally, we show in Fig.~\ref{fig:ldos} the local density of states (LDOS) of the interface before and after N-atom incorporation. The LDOS is calculated as
\begin{equation*}
\rho(z,E)=\sum_{i,k} \int |\Psi_{i,k}(x,y,z)|^2 \mathrm{d}x \mathrm{d}y \times N \mathrm{e}^{-\alpha(E-\varepsilon_{i,k})^2},
\end{equation*}
where $\varepsilon_{i,k}$ are the eigenvalues of the wavefunction with indexes $i$ and $k$ denoting the eigenstate and the $k$-point, respectively, $z$ is the coordinate of the plane where the LDOS is plotted, and $N(=2\sqrt{\frac{\pi}{\alpha}})$ is the normalization factor with $\alpha$ as the smearing factor. Here, $\alpha$ is set to 13.5~eV$^{-2}$. The LDOS at the interface is plotted on the plane, which includes the topmost C atom of the SiC substrate, while the plane for the bulk is chosen to be far from the interface so that the interactions between the SiO$_2$ and nitride layers are eliminated. The LDOS is shifted so that the center of the bandgap of the LDOS in the bulk corresponds to zero energy. It is found that the states arising from the CO bonds lie near the valence band edge of the bandgap in all the models with the CO bonds before N-atom incorporation. The effect of the CO bonds is significant at the C face; we confirm that this effect survives even when the thickness of the oxide layer is increased. By introducing the nitride layer between the SiC substrate and the SiO$_2$ layer, we find that the gap states at the topmost layer of the SiC side of the interface disappear, indicating that the N atoms incorporated by NO annealing can contribute to the reduction in the density of interface defects.

\section{Conclusion}
\label{sec:Conclusion}
The DFT calculations were carried out to investigate the electronic structures and stability of the 4H-SiC/SiO$_2$ interface after NO annealing using the assumption derived from the experiments\cite{JSurfSciNanotechnol_15_109, JApplPhys_97_074902} that the N atoms are fixed at the C atom site of the SiC. N-atom incorporation by NO annealing, in which a V$_\mathrm{Si}$ is formed and four C atoms adjacent to the V$_\mathrm{Si}$ are replaced by N atoms, is an exothermic reaction. Since the formation energy for the nitride layer growing along the $a$ face is the smallest even when the models containing SiO$_2$ are employed, the dependence of the areal N-atom density on the crystal plane is induced by the kinetic factors during the reaction rather than thermodynamic factors. In addition, in the case of the interface without the CO bonds, the nitride layer at the topmost layer of the interface is more stable than that at the second bilayer, resulting in the formation of the nitride layers at the topmost layer of the interface. We also investigated the interfaces in which the CO bonds remain after the thermal oxidation. Although the CO bonds at the interface prevent the generation of the nitride layer at the topmost layer of the SiC side, the CO bonds are removed before NO annealing because the reaction between the CO bonds and NO molecules is exothermic. From these results, it is reasonable to conclude that NO annealing forms the nitride layers immediately below the SiO$_2$ layers without any transition layers. LDOS analysis revealed that the states relevant to the CO bonds at the interface lie near the valence band edge of the bandgap in all the interface models. On the other hand, there are no gap states in the interface after N-atom incorporation. These results imply that NO annealing after the oxidation can contribute to the reduction in the density of interface defects by forming the nitride layer immediately below the SiO$_2$ layer of the interface.

\acknowledgments 
This work was partially financially supported by MEXT as part of the ``Program for Promoting Researches on the Supercomputer Fugaku'' (Quantum-Theory-Based Multiscale Simulations toward the Development of Next-Generation Energy-Saving Semiconductor Devices, JPMXP1020200205), JSPS KAKENHI (JP16H03865), Kurata Grants, and the Iwatani Naoji Foundation. The numerical calculations were carried out using the computer facilities of the Institute for Solid State Physics at The University of Tokyo, the Center for Computational Sciences at University of Tsukuba, and the supercomputer Fugaku provided by the RIKEN Center for Computational Science (Project ID: hp210170).

\end{document}